# Molecular dynamics of glycerol and glycerol-trehalose bioprotectant solutions nanoconfined in porous silicon.

R. Busselez[1], R. Lefort[1], M. Guendouz[2], B. Frick[3], O. Merdrignac-Conanec[4] and D. Morineau[1a]

[1] *Institute of Physics of Rennes, CNRS-University of Rennes 1, UMR 6251, F-35042 Rennes, France*

[2] *Laboratoire d'Optronique, FOTON, CNRS-University of Rennes 1, UMR 6082, F-22302 Lannion, France*

[3] *Institut Laue-Langevin, 6 rue Jules Horowitz, F-38042 Grenoble, France*

[4] *Sciences Chimiques de Rennes, CNRS-University of Rennes 1, UMR 6226, F-35042 Rennes, France*

**ABSTRACT**

Glycerol and trehalose-glycerol binary solutions are glass-forming liquids with remarkable bioprotectant properties. Incoherent quasielastic neutron scattering (QENS) is used to reveal the different effects of nanoconfinement and addition of trehalose on the molecular dynamics in the normal liquid and supercooled liquid phases, on a nanosecond timescale. Confinement has been realized in straight channels of diameter $D$=8 nm formed by porous silicon. It leads to a faster and more inhomogeneous relaxation dynamics deep in the liquid phase. This confinement effect remains at lower temperature where it affects the glassy dynamics. The glass transitions of the confined systems are shifted to low temperature with respect to the bulk ones. Adding trehalose tends to slow down the overall glassy dynamics and increases the non-exponential character of the structural relaxation. Unprecedented results are obtained for the binary bioprotectant solution, which exhibits an extremely non-Debye relaxation dynamics as a result of the combination of the effects of confinement and mixing of two constituents.

Monday, January 26, 2009

---

[a] Corresponding author : denis.morineau@uniuv-rennes1.fr

## 1. Introduction

The glass transition and the molecular dynamics of fluids can be markedly different under confinement in mesoporous materials. Although this fact is revealed by a large amount of experimental and numerical data,[1,2] these data also reveal numerous contradictory expressions of nanoconfinement effects. For instance the glass transition temperature ($T_g$) usually decreases in confinement but may also remains constant or increases.[1,3,4,5,6,7,8] Non-monotonic variations of $T_g$ as a function of the pore size of a series of mesostructured porous silicates have also been reported.[9,10] In some cases, confinement also leads to the observation of two distinct glass transition temperatures.[11,12,13,14] Although no conciliating viewpoint on these results has been reached yet, it is recognized that the nature of the confining medium, of the interaction between fluid and matrix as well as intermolecular interactions are essential issues. Key-parameters of interest include the topology of the confinement in terms of pore connectivity and dimensionality (isolated channels, spherical cavities, fractal or bi-continuous disordered structures), the nature of the porous matrix (smooth or hard confinement),[15] the chemical nature (hydrophilic, silanized) and the shape of the interface (rough or smooth).[1,2]

Beyond the simple determination of the glass transition temperature, a variety of experimental spectroscopies (quasielastic neutron scattering, dielectric spectroscopy, depolarized dynamic light scattering) and molecular simulation techniques have been used to investigate confinement effects on the molecular dynamics and especially the structural relaxation processes related to the glass transition.[5,10,11,12,13,16,17,18,19,20,21] They provide an extended accessible timescale (from a few picoseconds to the macroscopic limit of about a thousand seconds), which allows following the signatures of confinement as a function of temperature from the liquid phase down to the glass transition. Despite the persisting disparity of the observations (slowing down or speeding up of the relaxation dynamics), these results emphasize an apparently general feature of confined fluids: nanoconfinement magnifies the non-Debye character of the structural relaxation process. This feature is commonly related to dynamical heterogeneities for bulk glass-forming system.[22] In confinement, molecular dynamics simulations provide a microscopic insight into the molecular dynamics, which allows attributing the additional broadening of the distribution of relaxation times to a spatially highly inhomogeneous dynamics in the pore.[21]

Despite the number of interpretations proposed in literature, the problem of the molecular dynamics of nanoconfined fluids is still an open question. It certainly stresses the existence of many competing effects, the relative contributions of which depend on the characteristics of the system. Surface effects have been generally invoked and can be crucial according to the large surface-to-volume ratio of mesoporous materials. It is based on the introduction of a boundary condition, which is experienced by the molecules at the surface of the pore. This interfacial condition depends on the fluid-wall interaction and the roughness of the interface. It could either concern only the population of molecules localized at the interface (two-states situation) or propagate to the inner pore via a mechanism mediated by the extension of regions of dynamically correlated molecules.[11,12,13,16,21] Another major issue concerns the thermodynamic state of the confined phase, which may differ from the bulk one in terms of structure or density.[9,23] The structure of H-bonding fluids is likely to be more sensitive to confinement since they develop interfacial and intermediate range orders.[24,25] Differences in density can be induced by surface tension and also because the timescale for density equilibration through a liquid flow within the porous material exceeds the experimental one at low temperature.[5,9,26] Finally, finite size is expected to prevent the extension of any correlation length beyond the pore size. This potential effect is pertinent to supercooled liquids since the glassy dynamics is commonly associated to the growth of nanometric cooperative rearrangement regions or frustration limited domains.[27,28] However, a definite signature of such intrinsic geometrical confinement effects seems elusive because of the complex entanglement with the above mentioned dominant effects.

Although there is still a lot of discussion even in the case of pure low-molecular weight simple glass-forming liquids, we are convinced that it is timely to extend the problem of nanoconfinement to more complex fluids. The latter are more relevant to different domains of technological or biological interest and also they can pose new questions for fundamental research. One example concerns mesogenic fluids, the phase transition and glass-like dynamics of which have been shown to be sensitive to confinement induced quenched disorder effects and low dimensionality.[29,30,31] Another example of such fluids is bioprotectant solutions,[32,33] which are used to prevent proteins and biological membranes from irreversible damage under drastic stress environmental conditions (e.g. drying or freezing conditions). They are often binary systems with strong and selective H-bond interactions (like sugar-water or sugar-polyalcohol solutions), which can lead to an unusual behavior under nanoconfinement. For instance, concentration fluctuations can be amplified by a possible asymmetric affinity of one particular constituent with the pore surface. In the case of glass-forming constituents with very different $T_g$, the relaxation dynamics of the mixture in the liquid phase is expected to cover a very broad range of characteristic times. As an indirect sign, the scrutiny of the resulting $T_g$ of the bulk solution versus concentration can reveal its non regularity (departure from the Gordon-Taylor law).[34,35] Here again, these effects are expected to be strongly affected by nanoconfinement.

The aim of this paper is indeed to identify some generic features related to confinement and interfacial effects on the molecular dynamics of bioprotectant solutions. This investigation is performed in controlled model conditions by using porous silicon material as a confinement matrix. It has been shown to provide unique experimental conditions to study nanoconfined complex fluids due to the low dimensionality of its macroscopically oriented pores.[30] Glycerol/Trehalose binary solutions are chosen as a model bioprotective fluid. These solutions are formed with two glass-forming liquids with very different glass transition temperatures, as measured by differential scanning calorimetry for the two pure components trehalose $T_g$=393 K[36,37] and glycerol $T_g$=190 K.[7] The addition of 20 % (Wt) trehalose to glycerol leads to an upward shift of Tg of about 13 K, which is almost twice smaller than what is predicted by assuming a regular solution model.[34,38] Such a discrepancy probably reflects effects of non-idealities, which have been already reported for binary systems with different molecular size and specific H-bond interactions.[39,40] They have been recently highlighted by experiments and simulations for their optimal biopreservation properties.[41,42] Incoherent quasielastic backscattering experiments are performed on fully hydrogenated pure glycerol and a glycerol-trehalose binary solution (80%:20% Wt) both in bulk and confined in nanoporous silicon layers. The effects of adding trehalose and confinement effects on the structural relaxation dynamics in the liquid state are discussed from the analysis of the dynamic structure factor. The consequences on the dynamical arrest of the relaxation dynamics on approaching the glass transition are revealed by the analysis of elastic fixed window temperature scans on cooling.

## 2. Experimental procedure

### 2.1. Samples

Porous silicon matrices were made from a crystalline silicon substrate using an electrochemical anodization process in a HF electrolyte solution. The conditions of this process control properties such as porosity, thickness and pore size. Anodization of heavily *p*-doped (100) oriented silicon leads to highly anisotropic pores running perpendicular to the surface wafer (called columnar form of the PSi). These samples were electrochemically etched with a current density of 50 mA.cm$^{-2}$ in a solution composed of HF, H$_2$O and ethanol (2:1:2) according to previous studies.[43] Scanning electron microscopy (SEM) measurements provide interesting information about the shape of the porosity. It forms a parallel arrangement of not-connected channels of length 200 $\mu$m as shown in Fig. 1.

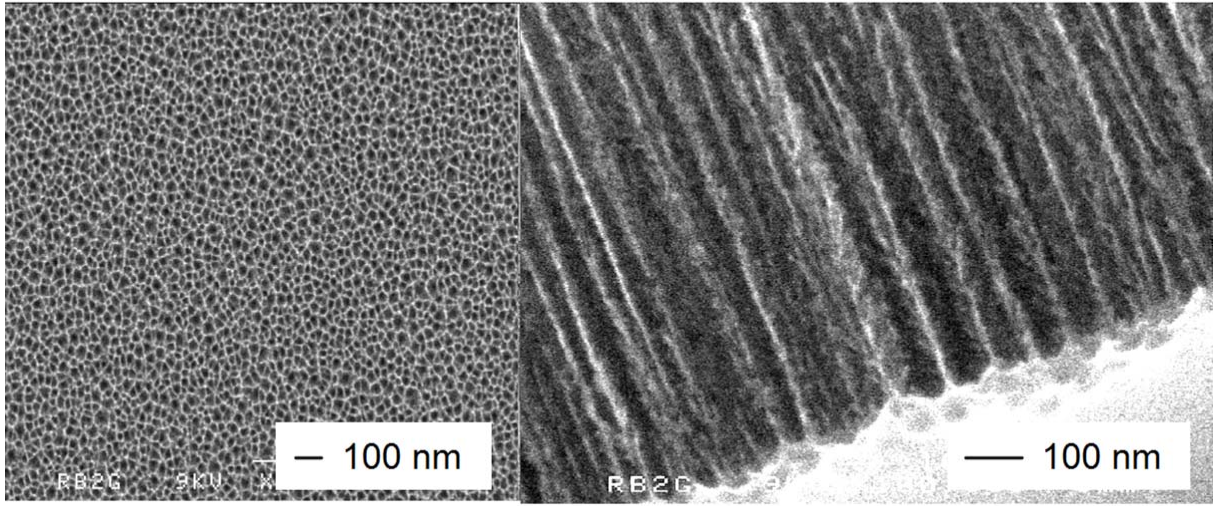

**FIG. 1:** Scanning electron micrographs of the porous silicon film. (a) top view at low magnification showing the 30 $\mu$m thick porous layer attached to the silicon substrate and (b) side view at higher magnification.

A noticeable feature of PSi, which is at variance to other unidirectional pores such as MCM-41 or SBA-15, is the rough dendritic structure of the inner pore surface.[44,45] The latter has been recently shown to introduce strong disorder effects, which influence phase transitions such as capillary condensation or nematic to smectic transition.[30,46] SEM has shown to provide a rather poor estimation of the pore size. It can be explained by the low spatial resolution of this technique, which is even deteriorated by the low conductivity of PSi, and by the fact that only the pore opening is checked and not the inside porous layer. Therefore, nitrogen adsorption isotherms have been preferred to characterize the pore size. The analysis of the onset of capillary condensation in the nitrogen adsorption/desorption curves with a model based on the Kelvin equation provides an average value of the pore diameter of 8 nm, as shown in Fig. 2.

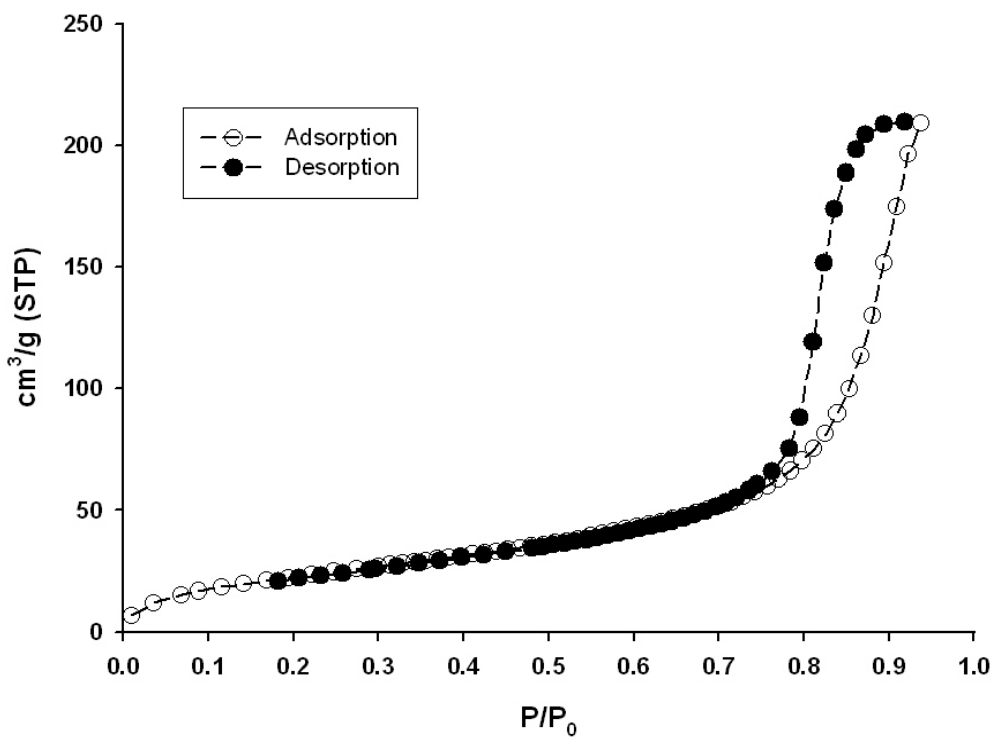

**FIG. 2:** Adsorption/desorption isotherms of nitrogen at 77 K in porous silicon films.

The strongly hydrophobic character of the inner surface of freshly prepared PSi was minimized by treating the sample with hydrogen peroxide ($H_2O_2$). $H_2O_2$ leads to a slow oxidation of the proximity of the inner pore surfaces.[47] A droplet of $H_2O_2$ was spread on the porous surface and the sample was placed in a vacuum chamber. Capillary impregnation of $H_2O_2$ in the entire porosity was achieved under primary vacuum at 293 K. Atmospheric pressure was then restored in the chamber and the sample maintained for 15 minutes for the oxidation to proceed. PSi wafers were dried under vacuum (4. $10^{-2}$ mbar) at 373 K for 24 h and stored in the vacuum chamber until used.

Fully hydrogenated glycerol ($C_3O_3H_8$, 99%) and anhydrous trehalose ($C_{12}H_{22}O_{11}$, 99%) were purchased from Sigma-Aldrich and Merck, respectively. A glycerol-trehalose binary solution (80%:20% Wt) was prepared by adding the appropriate mass of crystalline trehalose to glycerol. The mixture was maintained in a hermetically closed flask, then heated up to 333

K and stirred for about 4 hours until the solution was optically transparent as a consequence of full dissolution of trehalose. Just before the execution of neutron scattering experiments, pure glycerol and trehalose-glycerol were filled in PSi samples by capillary impregnation from the liquid phase in a vacuum chamber. The complete loading was achieved under vapor pressure and at a temperature of 333 K. The filled porous membranes were taken out from the vacuum chamber, the excess of liquid was removed from the wafer surface by wiping the samples with filtration papers and the samples were placed immediately in hermetically closed neutron scattering aluminum cells.

*2.2. Quasielastic neutron scattering*

Quasielastic neutron scattering experiments were carried out with fully hydrogenated samples using the high resolution back scattering spectrometer (BS) IN16 at the Institut Laue Langevin (Grenoble).[48] For the two molecular fluids, the contribution from the incoherent cross section corresponds to more than 93% of their total scattering cross section. It means that, after appropriate subtraction of the intensity arising from the sample environment and PSi, the contribution to the measured scattering intensity from coherent scattering can be neglected within a good approximation for the four samples. Moreover, 85% of the incoherent scattering of the glycerol-trehalose binary solution is arising from glycerol molecules. This means that despite a predominant contribution from glycerol to the incoherent scattering, results obtained for the mixture are composite functions, which reflect the dynamics of both constituents of the solution. A standard configuration of the IN16 spectrometer was chosen with Si(111) monochromator and analyzers in backscattering geometry, which corresponds to an incident wavelength of 6.271 Å and results in an energy resolution (FWHM) of 0.9 $\mu$eV. The energy range was 15 $\mu$eV with a range of transfer of momentum $Q$ between 0.2 Å$^{-1}$ and 1.9 Å$^{-1}$.

PSi layers were placed in cylindrical aluminium cells. The selected orientation of the cell leads to a pore axis angle with respect to the incident neutron beam of about $\omega=45°$. Anisotropic dynamics is not expected for this system. The choice of this angle of incidence essentially guarantees that self-attenuation effects, which occur in the direction parallel to the PSi wafer surface, are significant only for an angle of diffraction of about $2\theta \approx 135°$. Such large values are outside our range of interest. A cryofurnace was used in order to regulate the sample temperature in a range from 10 K to 350 K.

Elastic fixed window scans (EFWSs, i.e. with Doppler device stopped) were carried out in a temperature range from 10 K to 350 K. EFWSs have been performed both on cooling with typical ramps of 0.5 K.min$^{-1}$ from 350 K down to 130 K and with 2 K.min$^{-1}$ below. Data accumulation was performed during the continuous temperature scan integrating over intervals of 1 to 2 K. Quasielastic scattering functions $S(Q,\omega)$ were acquired at 310 K for the four different samples. The spectrometer resolution function $R(Q,\omega)$ was measured on each sample at 10 K. Standard data corrections were applied to EFWSs and scattering functions using conventional programs provided at ILL (*sqw*). The intensity was corrected for the contribution arising from the empty sample (empty cell and empty PSi) and thereafter normalized to the lowest temperature intensity. These background contributions appeared as purely elastic signals on IN16. There were extremely small for the bulk samples. For the confined systems, the contribution arising for the silicon wafers accounted for 25% to 35% of the maximum intensity at the highest intensity. The fitting of scattering functions $S(Q,\omega)$ in the frequency domain was carried out using the programs provided by the LLB (*quensh*).[49]

### 3. Results and discussion

*3.1. Molecular dynamics in the high temperature liquid phase.*

Quasielastic neutron scattering experiments have been used to investigate confinement effects on the molecular dynamics on a typical timescale from 0.1 to 2 nanoseconds. The scattering intensity from the fully hydrogenated molecules essentially corresponds to the incoherent scattering function $S_{inc}(Q,\omega)$, which allows to probe the self-correlation of the hydrogen motion. This function is commonly approximated by Eq. 1 :

$$S_{inc}(Q,\omega) = \exp\left(-\frac{\langle u^2 \rangle Q^2}{3}\right)\left[A(Q)\delta(\omega) + (1-A(Q))S_{quasi}(Q,\omega)\right] \qquad (1)$$

In this expression, $A(Q)$ stands for the elastic incoherent structure factor coming from restricted motions, $S_{quasi}(Q,\omega)$ is the quasielastic response of slow molecular relaxations, and the contribution of the fast vibrational modes is approximated by the Debye-Waller factor expressed as an overall intensity loss expressed by the vibrational mean square displacement $\langle u^2 \rangle$.[50]

The incoherent scattering function has been measured for the four systems at $T$=310 K, in order to characterize the liquid dynamics above the glass transition temperature, as shown in Fig. 3. The incoherent scattering function appears essentially as a quasielastic line, which is broader than the elastic resolution ($A(Q)$=0, $S_{inc}(Q,\omega) \propto S_{quasi}(Q,\omega)$). The absence of any pure elastic contribution proves the loss of correlation of the protons motion on the time scale of the experiment.[50] This non-localized character of the motion of the molecules most probably implies that translational diffusion processes occur on the time and length scales of the experiment. Comparing the line width of the scattering functions, one can also deduce the qualitative conclusion that the relaxation dynamics of the pure glycerol samples is faster at 310 K than for the binary solution samples. Nevertheless, a more quantitative analysis of the shape of the quasielastic lines is required to draw more detailed conclusions.

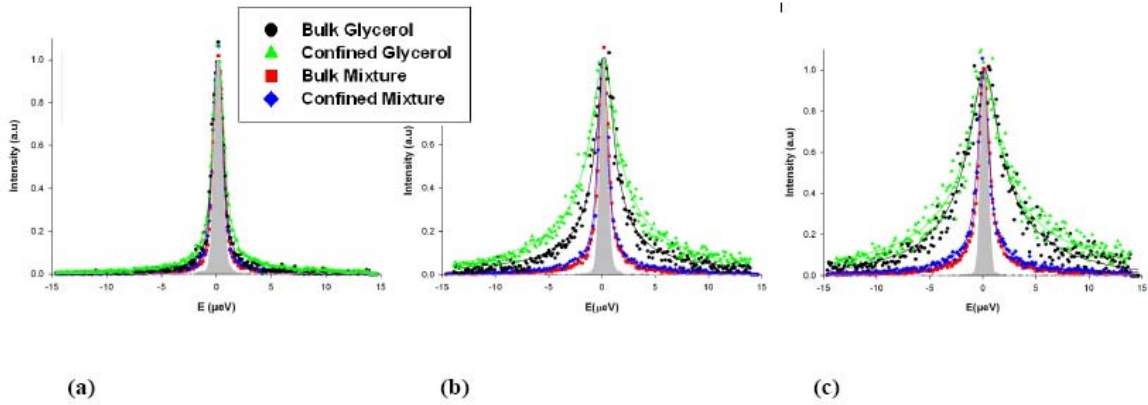

**FIG. 3:** Incoherent quasielastic spectra of glycerol and the glycerol-trehalose solution in bulk and confined in porous silicon layers measured at *T*=310 K on the high resolution backscattering spectrometer IN16. The intensity is corrected for empty sample contribution and normalised to maximum intensity. The displayed curves have been scaled to one at the maximum intensity for clarity. The resolution function of the apparatus is displayed as a filled gray shape. The solid lines are best fits of the data using a stretched exponential (see text). The graphs correspond to three different values of the transfer of momentum, (a) $Q$=0.54 Å$^{-1}$, (b) $Q$=0.96 Å$^{-1}$ and (c) $Q$=1.33 Å$^{-1}$.

The scattering functions have been fitted at each value of the transfer of momentum $Q$ by the Fourier transform of a Kohlrausch function and a weak constant background $f(Q)$ convoluted with the experimental resolution, using the *quensh* software, according to equation (2). The obtained values of $b(Q)$ are almost within experimental uncertainties, with no systematic variations nor actual physical meaning. However, their use improves the quality of the fit by accounting for statistical errors in the evaluation of empty container background.

$$S_{quasi}(Q,\omega) = f(Q) R(\omega) \otimes FT\left\{\exp\left(-\left(\frac{t}{\tau_K(Q)}\right)^{\beta_k}\right)\right\} + b(Q) \qquad (2)$$

This stretched-exponential law is recognized as a convenient empirical way to reproduce the relaxation functions of glass-forming materials. This single function usually better reproduces the experimental data than a discrete combination of different modes (translation, rotation…), which are most probably coupled in the case of glass-forming liquids.[51,52] $\tau_K$ stands for the characteristic relaxation time, whereas $\beta_K$ reflects the non-Debye character of the relaxation function and the average relaxation time is $\langle \tau \rangle = \dfrac{\tau_K}{\beta_K} \Gamma\left(\dfrac{1}{\beta_K}\right)$.

This procedure provides good fits of the data for the four systems as shown as solid lines in Fig. 3. In a first step, both $\tau_K$ and $\beta_K$ values were fitted for each $Q$ values. Since no significant and systematic $Q$ dependence of the $\beta_K$ exponents were observed, in a second step, they have been fixed for each sample to an average value and only $\tau_K(Q)$ values were refined.

The value of $\beta_K$ exponents for pure glycerol is about 0.6. Different values for the $\beta_K$ exponent have been reported for pure glycerol. As reported in ref. 53, the shape of the relaxation function may depend on the experimental method used to probe the liquid dynamics. In fact, the value in our study is in full agreement with previous quasielastic neutron scattering experiments performed on a larger temperature range and using time-temperature superposition principle (0.58±0.02).[51,54] Interestingly, $\beta_K$ differs significantly for the binary solution ($\beta_K$= 0.5 for the bulk binary solution) and is also reduced by confinement, being equal to $\beta_K$=0.5 and $\beta_K$=0.4 for the pure confined glycerol and the confined binary solution, respectively. The exact values of $\beta_K$ should be taken with some care, since they result from a fit at different $Q$-values but at a unique temperature and covering a limited frequency (or time) scale of about one decade only. However, the difference between the $\beta_K$ exponents for the different samples is so large, that uncertainties coming for the fitting procedure could not affect the qualitative conclusions concerning trehalose and confinement effects on the dynamics.

It is well-known that the stretched exponential relaxation function may be expressed within two different scenarios, corresponding to the limiting cases of heterogeneous and homogeneous dynamics.[22] The homogeneous scenario assumes that the relaxation process is intrinsically of non-Debye type. On the contrary, the heterogeneous scenario relates stretching to the existence of dynamical heterogeneities of given typical size and life-time. Each individual heterogeneity may relax according to a simple Debye mode, with a different relaxation time, leading on average to a broadening of the distribution of relaxation times.

The marked decrease of the $\beta_K$ exponent in the binary system as compared to pure glycerol is most probably a direct consequence of a more heterogeneous dynamics. The bulk solution is composed of two glass-forming molecules with very different glass transition temperatures. The calorimetric glass transition of the solution depends strongly on the concentration.[38] The local molecular dynamics can be affected likewise by the existence of concentration fluctuations at the nanometer scale, which are inherent to binary solutions. Moreover, this tendency to form spatial heterogeneities is amplified by the H-bonding character of the two molecules, which usually favors the formation of transient networks and clusters of mesoscopic size.[55] Indeed, trehalose and homologous disaccharide aqueous solutions are known to develop structural inhomogeneities, in terms of clusters and transient H-bond networks.[56] They are probably responsible for the major part of the broadening of the distribution of times of the structural relaxation with respect to neat glycerol.

Confinement also leads to a reduction of the $\beta_K$ exponent. This observation is in agreement with previous QENS experiments and molecular dynamics simulations of other nanoconfined glass-forming liquids.[19,21] This increase of the non-Debye character of the relaxation is likely to be attributed to the formation of spatial heterogeneities within the pore, which dynamical parameters depend on the distance to the pore surface. There are indeed an increasing number of experimental and numerical evidences that the dynamics of a confined fluid is spatially heterogeneous.[16,19,21,57,58,59] The case of the confined trehalose-glycerol solution is very striking, leading to a very small $\beta_K$ exponent ($\beta_K$ =0.4). For a binary glass-forming liquid, it is not excluded that dynamical heterogeneities can be amplified by non-homogeneous concentration distribution profile across the pore diameter.

Figure 4 shows the $Q$-dependence of the average relaxation rate $1/\langle\tau\rangle$. This quantity is preferred to $\tau_K$, since it is much less sensitive to statistical uncertainties induced by numerical correlations between $\tau_K$ and $\beta_K$ in the fitting procedure. Power law fits of $\langle\tau\rangle$ give $\langle\tau\rangle \propto Q^{-\nu}$ with an exponent close to $\nu=2$. A value of $\nu$ slightly larger than 2, is also encountered for another polyalcohol, ethylene glycol.[60]

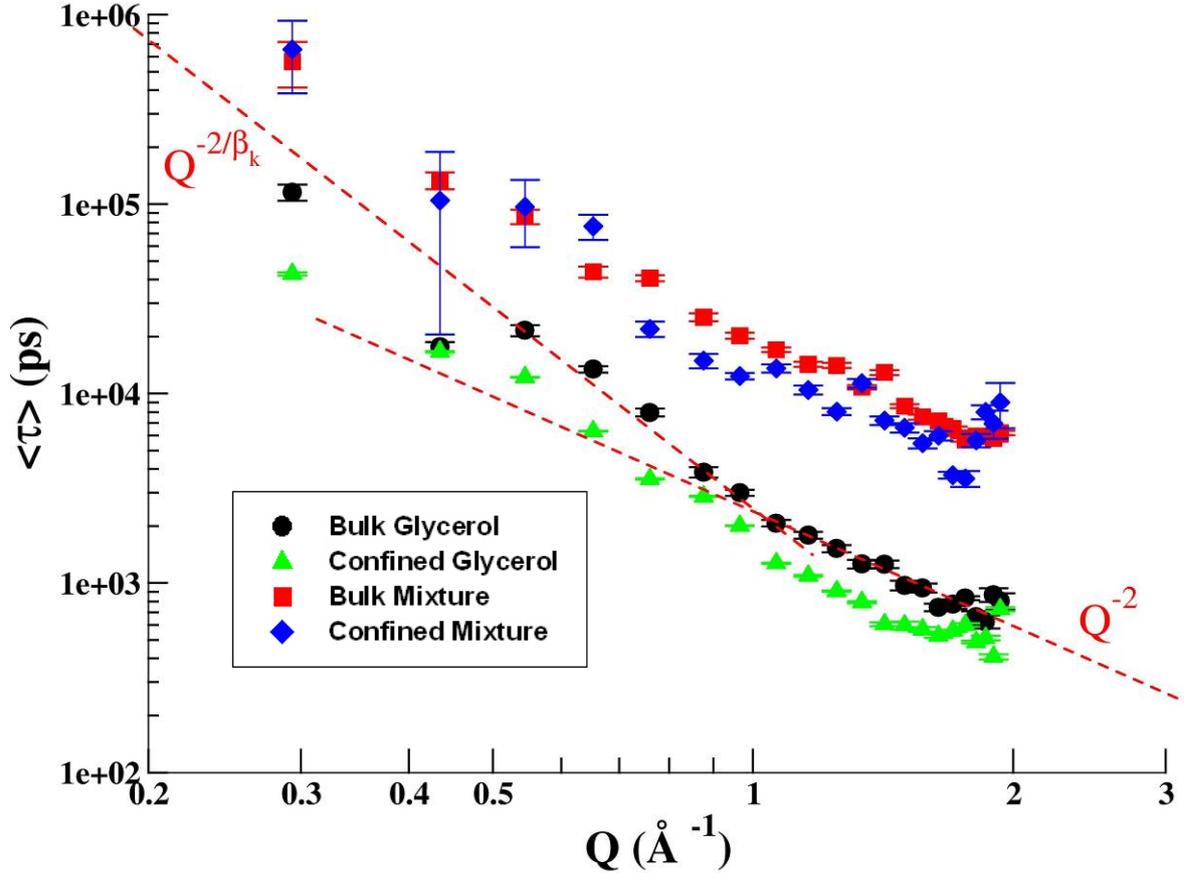

**FIG. 4:** Average relaxation time $\langle\tau\rangle$ as a function of the transfer of momentum $Q$ obtained from the incoherent quasielastic spectra at $T=310$ K for pure glycerol and the glycerol-trehalose solution in bulk and confined in porous silicon. Dashed lines superimposed on bulk glycerol data emphasize the cross-over between two different power law variations.

In the case of pure glycerol a departure from this single power law is observed at low $Q$ ($Q<1$ Å$^{-1}$), leading to a steeper $Q$-dependence. This feature has already been reported for bulk glycerol and more usually in glass-forming polymers.[54,61,62] A dispersion relation with $\nu=2$ in the case of a stretching exponent $\beta_K$ smaller than 1 means that the Gaussian approximation of the shape of the van Hove self correlation function does not hold anymore. In fact, our observation supports the idea of a possible non-Gaussian to Gaussian cross-over around the location of the main diffraction peak, as already discussed for pure glycerol.[54,62] It highlights the coupling of the self-diffusion to the structural relaxation on the time and length scales probed by QENS. The analysis of this feature, including the case of a binary mixture and confined systems goes beyond the scope of the present paper, but will be discussed in a future study combining ongoing molecular dynamics simulations.[38]

In order to address the influence of confinement on the molecular dynamics of the investigated liquids, we have compared the slopes of the dispersion curves in the $Q$-range where the $Q^{-2}$ power law holds ($Q>1$ Å$^{-1}$). The incoherent intermediate scattering function relaxes mainly via the translational diffusion of the protons, with a reduced amplitude due to fast relaxation and vibrational processes. Assuming a model based on an inhomogeneous distribution of normal diffusion processes, it can be expressed as follows: [63]

$$S_{quasi}(Q,t) = f(Q)\exp\left(-\left(D_w Q^2 t\right)^{\beta_k}\right) \qquad (3).$$

From the dispersion curve, one can obtain an average diffusion coefficient $<D>=1/(<\tau>Q^2)$, as already described for other liquids, such as confined water.[64] The values of $<D>$ at the temperature $T=310$ K are reported for the four samples in Table 1. The value obtained for bulk glycerol is $<D>=5.10^{-8}$ cm$^2$.s$^{-1}$, in quantitative agreement with a recent NMR study.[65] The addition of 20 percents trehalose considerably reduces the average diffusion coefficient of the solution, which is ten times smaller than for pure glycerol. Moreover, the diffusion coefficients under confinement are systematically about 1.5 larger than for the bulk liquids (see Table 1 and also Fig. 4) both for the pure and the binary solution. This result demonstrates that the microscopic molecular dynamics is already affected by confinement in the normal liquid phase above the melting point.

**Table 1**

Average diffusion coefficient and Kohlrausch stretching exponent obtained from the incoherent quasielastic neutron backscattering function at a temperature $T = 310$ K.

|  | Bulk glycerol | Confined glycerol | Bulk trehalose-glycerol | Confined trehalose-glycerol |
|---|---|---|---|---|
| $\beta_K$ | 0.6 | 0.5 | 0.5 | 0.4 |
| $<D>$ ($10^{-8}$cm$^2$.s$^{-1}$) | 5.0 | 7.6 | 0.52 | 0.77 |

*3.2. Variable temperature fixed window scans*

In order to monitor the molecular dynamics from the liquid to the glassy state, elastic fixed window scans (EFWS) have been acquired during cooling from 350 K to 10 K, as shown in Fig. 5. In a EFWS, the measured intensity is in principle $S_{inc}(Q,\omega=0)$ (see equation 1). It actually integrates all purely elastic ($A(Q)$) or quasielastic contributions slower than the energy resolution defining the "fixed window" (c.a. 0.9 $\mu$eV FWHM). At high temperature (above 350 K), the elastic intensity is below the detection level for bulk and confined glycerol and extremely small for the bulk and confined binary solution. This result is expected for a liquid for which translational diffusion completes within the experimental timescale, giving rise to a quasielastic line that fully exceeds the elastic resolution, as shown in the precedent part. The elastic intensity gradually increases on cooling to reach a value of about 0.8 at 200 K. This feature is common to glass-forming systems. It most probably reflects the progressive slowdown of the translation diffusion and possibly rotational molecular dynamics, which leads to a sharpening of the associated quasielastic scattering. The contribution of this quasielastic scattering to the fixed energy window intensity increases until the quasielastic lines are not broad enough to be discriminated from a true elastic peak within the experimental resolution. Below this temperature of onset of structural relaxation processes (glass transition on the time scale of the neutron scattering experiment), only localized degrees of freedom of the glass phase (conformation fluctuation and Debye-Waller factor, comprising inter- and intramolecular vibrational modes) result in a remaining (but weaker) temperature dependence of the elastic component.

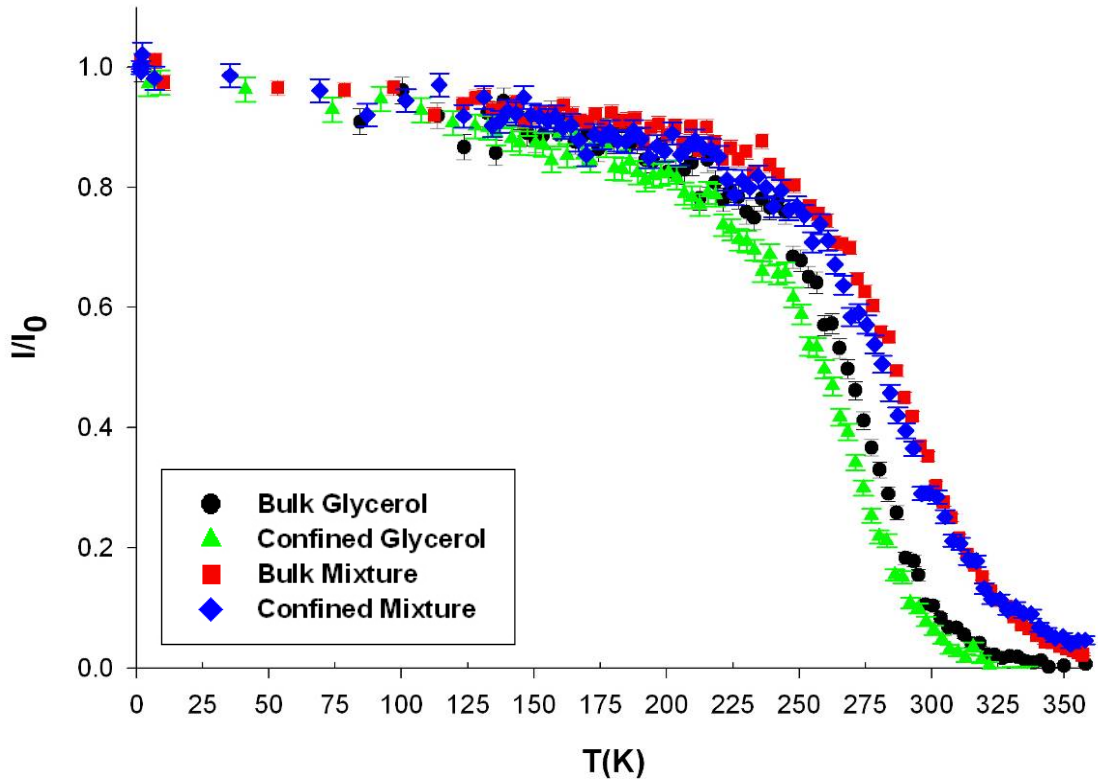

**FIG. 5:** Fixed window elastic scattering intensity of glycerol and the glycerol-trehalose solution in bulk and confined in porous silicon layer obtained during a scan on cooling for transfer of momentum $Q=1.33$ Å$^{-1}$. The intensity is corrected for empty sample contribution and normalised at the lowest temperature.

A comparison between the EFWS displayed in Fig. 5 for the four different systems reveals some distinct effects on the relaxation dynamics of glycerol solutions on the nanosecond timescale of both confinement and of the addition of trehalose. The elastic intensity of the bulk trehalose-glycerol solution qualitatively resembles the bulk glycerol one, except for an overall shift to higher temperature of about 20 K, when measured at half the maximum elastic intensity. This overall slowing down of the relaxation dynamics of the binary solution on increasing the amount of trehalose is mostly related to the change in the glass transition temperature. Experiments with specifically H/D labeled binary systems could partly disentangle the different processes involving glycerol and trehalose molecules, which contribute to the overall slowdown of the relaxation decay probed for fully hydrogenated solutions.

Conversely, confinement leads to an apparent shift to lower temperature of the elastic intensity of about $\Delta T$= -7 ±2 K when measured a half maximum intensity. Interestingly, the confinement effect cannot be described by a simple temperature shift. Indeed, the slope of the elastic intensity as a function of temperature is smaller in confinement than in bulk in the temperature region where the time scale of quasielastic relaxation processes crosses the instrumental resolution (i.e. between 300 K and 250 K for glycerol and between 330 K and 270 K for the binary solution). This feature is observed for pure glycerol and for binary solution. As a consequence, the difference of the elastic intensity between bulk and confined pure glycerol keeps increasing on decreasing temperature from about 300 K to 250 K and approaching the glass transition. The situation of the trehalose-glycerol solution is even more striking, since the two elastic intensities cross at about 325 K. This observation is of central interest to understand confinement effect. A broadening of the transition region could reflect a broader distribution of relaxation times related to a higher degree of heterogeneity of the dynamics in confinement. In addition, it may arise from an increasing deviation between the dynamics of the bulk and confined liquid during cooling on approaching the glass transition. It is also remarkable that the elastic intensity vanishes or is very small at 350 K for the two confined fluids (i.e. 50-60 K above $T_g$ of bulk glycerol). This differs from previous neutron spin echo and neutron backscattering experiments on confined liquids, which show a remaining elastic intensity well above the glass transition: more than 20% of elastic intensity 150 K above $T_g$ for the glass-former OTP in 7 nm SBA-15 porous silicates and 20% of elastic intensity 30 K above the clarification point for a mesogenic liquid 8CB in PSi.[20,31] This remaining apparent elastic contribution is commonly attributed to a fraction of the confined liquid, most probably located near the solid interface, which is immobile on the time scale of the experiment. Such surface induced blocking effect on the molecular dynamics is obviously of smaller importance in our case, probably because of a smaller surface to volume ratio or a different nature of the surface interaction.

*3.3. Low temperature mean square displacement*

At low temperature, the quasielastic lines due to relaxation processes are not broad enough to be discriminated from a true elastic peak within the experimental resolution. Vibrational modes lead to an inelastic contribution, which is outside the energy range covered by BS. They also lead to a reduction of the elastic intensity, which allows one to measure the mean square displacement (MSD) $<u^2>$ of vibrational modes from a linear regression of the logarithm of the elastic intensity as a function of $Q^2$ on a $Q$-range from 0.7 Å$^{-1}$ to 1.9 Å$^{-1}$ according to

$$S_{inc}(Q,\omega,T)/S_{inc}(Q,\omega,10K) = \delta(\omega)\exp\left(-\frac{\langle u^2 \rangle(T)Q^2}{3}\right). \qquad (4)$$

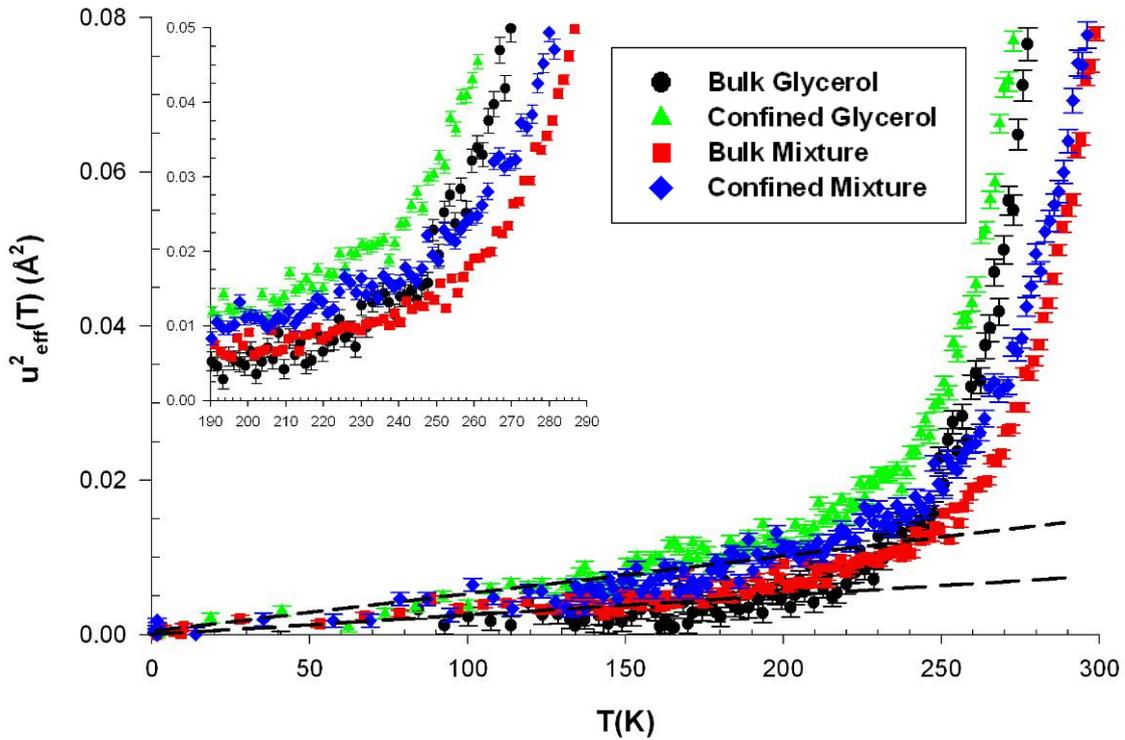

**FIG. 6:** Mean square displacement of glycerol and the glycerol-trehalose solution in bulk and confined in porous silicon layer obtained from incoherent neutron backscattering on cooling. Inset : magnification of the temperature region of the glass transition.

Figure 6 shows the measured mean-square displacement as a function of temperature. The values obtained for bulk glycerol are in quantitative agreement with previously published results.[51] In the temperature range from 10 K to 200 K, $<u^2>$ increases linearly with $T$, which is in agreement with a harmonic approximation of the vibrational dynamics of the glass. Above this temperature, the MSD increases more rapidly indicating the sequential onset of vibrational modes of larger amplitude, fast relaxation and the main structural relaxation. In the present case, these different modes show up in overlapping temperature ranges. Their different contributions to the MSD are hardly disentangled without any further data modeling, such as the self-distribution function (SDF) procedure.[66] This is at variance to some polymeric systems, which MSD exhibit well-defined low temperature modes related to side group dynamics and decoupled from the alpha relaxation.[67] The rise of the MSD shows up in a temperature region around the calorimetric glass transition temperature ($T_g$=190 K), as already observed for many other glass-forming molecular systems. These findings suggest that the fast dynamics probed by the MSD also senses the glass transition as other quantities such as enthalpy and specific volume, as discussed in ref. 68. In the case of the bulk binary solution, a similar behavior is observed but the deviation from the harmonic behavior occurs at higher temperature (about 210 K), in agreement with previously discussed EFWS. A similar cross-over is observed for the two confined systems, which suggest that the glass transition is also probed in confinement from the variation of MSD. Similar signatures of the glass transition are usually reported for confined molecular liquids and polymers,[67] although this upturn is not universally related to Tg as shown by the work of Soles, et al. on nanoconfined polycarbonate thin films.[69]

For both pure glycerol and the trehalose-glycerol solution, confinement results in an increase of the MSD in the glassy and liquid states. In the low temperature region, this suggests a weaker stiffness for the confined vitreous phase whereas in the high temperature region, the larger amplitude movements mostly result from the lower glass transition temperature of the confined phase as compared to the bulk. Here again, such a behavior does not seem to be universal to confined glassforming systems, and reductions of the MSD have been also reported for polymer thin films in the glassy state.[69]

### 4. Summary and conclusions

We have investigated the dynamics of pure glycerol and a trehalose-glycerol binary solution in bulk and confined situations. The incoherent dynamic structure factor measured by QENS has provided information on the proton self-motion in the liquid phase well above *Tg* (at *T*=310 K), corresponding to a typical time window of about 0.1 to 2 ns, which also corresponds to the liquid structural relaxation timescale at this temperature. EFWS have allowed measuring the molecular relaxation dynamics on a temperature range corresponding to the liquid and supercooled liquid. The vibrational and sub-nanosecond fast relaxation dynamics have been described by the mean square displacement $<u^2>$ in the glass (below 200 K) and in the region of the glass transition temperature.

It is noteworthy that the overall measurements provide a same qualitative depiction of trehalose and confinement effects on the dynamics of the glycerol solution. This is also in agreement with ongoing NMR and dielectric spectroscopy studies on the same materials.[Erreur ! Signet non défini.,70]

Trehalose significantly slows down the dynamics of the glycerol solution. At $T$=310 K in the liquid phase, it corresponds to a decrease of the average self diffusion coefficient by a factor of ten. This effect persists on decreasing temperature and affects the glassy dynamics, as reflected by a significant shift to high temperature of the EFWS ($\Delta T_g \approx 20$ K). This confirms the so-called 'switching-off' effect of trehalose, which has been related to its exceptional biopreservative efficiency.[32,33,41,42] The second observation is a smearing out of the relaxation steps. It is characterized by an increased non-Debye character of the incoherent intermediate scattering function in the liquid phase. It supports the idea that trehalose in glycerol promotes dynamical inhomogeneities. This is generally awaited for a binary system composed by molecules, which have different characteristic relaxation modes and most specifically for disaccharides bioprotectant solutions, which are known to develop structural inhomogeneities, in terms of clusters and transient H-bond networks.[56]

In addition, we have investigated the same two liquids under confinement using PSi nanochannels. At $T$ = 310 K, in the liquid phase, we have observed a slight modification of the incoherent quasielastic neutron scattering functions, which are broader and more stretched as compared to the same bulk systems ($\beta_K$ stretching exponent as small as $\beta_K$ = 0.4 for the confined binary solution). These effects have been quantified assuming a model of an inhomogeneous distribution of diffusion coefficients, which are coupled to the structural relaxation in the time and lengthscale of this study. The average diffusion coefficient is found systematically larger in confinement by a factor of 1.5 as compared to bulk value.

On decreasing temperature, fixed window scans confirm a deviation of the relaxation dynamics of the confined systems from the bulk one, which seems to increase on approaching the glass transition. It results in a substantial shift (estimated to about $\Delta T$ = -7 K) of the temperature where the MSD $<u^2>$ departs from its essentially vibrational behavior in the glassy state.

The broadening of the intermediate scattering function at $T$ = 310 K in the liquid state proves that confinement amplified the inhomogeneous character of the liquid dynamics. This is an evidence for the existence of non-equivalent dynamical environments within the pore. We note that this feature is likely to be amplified for a multi-constituent system with non-symmetric fluid-wall interactions.

These overall results (e.g. change of the average relaxation time and broadening of the relaxation distribution) agree with ongoing solid-state NMR and dielectric experiments on the same systems.[22,70] Such features have been frequently reported for other systems and often discussed with respect to some predictions about finite size effects. However, we note that these observations are not systematic among different systems and a unifying picture is certainly missing.[1,2,4,6,7]

It is also possible that the observed confinement effects on the glassy dynamics result from a modification of some other physical properties of the confined liquid (such as density, rheology or H-bonds induced intermediated range order), which indirectly affects the structural relaxation.[5,9,26] Very recently, there has been a growing number of evidences of long-lived mesoscopic dynamical heterogeneities in bulk liquid glycerol and long-time long-range relaxation processes in glycerol-aerosils dispersions well above $T_g$, which suggest that confinement may lead the confined glass-forming liquid to a frustrated state with different physical (static or dynamical) properties.[71,72,73]

For pure glycerol, our results emphasize significant confinement effects on the structural relaxation deep in the liquid phase as well as on the glassy dynamics on cooling, which relation to bulk mesoscopic dynamical heterogeneities is to be unraveled. We have also reported unprecedented manifestation of an amplified non-Debye character of the relaxation function for the binary bioprotectant solution that goes beyond the case of pure glycerol. Molecular simulations are required in order to scrutinize potential mesoscopic inhomogeneities related to concentration fluctuations and specific interfacial interactions.

**Acknowledgements**

We thank R. Pelster and P. Huber from the University of Saarbrücken for fruitful discussions and J. M. Zanotti from LLB Saclay for making available the software package *quensh* for neutron data modeling. R. B. has benefitted from a doctoral research grant from the *Brittany Region*. Financial supports from the *Centre de Compétence C'Nano Nord-Ouest* and *Rennes Metropole* are expressly acknowledged.

**REFERENCES**

[1] C. Alba-Simionesco, B. Coasne, G. Dosseh, G. Dudziak, K. E. Gubbins, R. Radhakrishnan, and M. Sliwinska-Bartkowiak, J. Phys.: Condens. Matter **18**, R15 (2006).

[2] M. Alcoutlabi and G. B. McKenna, J. Phys.: Cond. Mat. **17,** R461 (2005).

[3] C. L. Jackson and G. B. McKenna J. Non-Cryst. Solids **131–133** 221 (1991)

[4] J. Zhang, G. Liu, and J. Jonas, J. Phys. Chem. **96**, 3478 (1992).

[5] A. Patkowski, T. Ruths and E.W. Fischer, Phys. Rev. E **67**, 021501 (2003).

[6] O. Trofymluk, A. A. Levchenko, and A. Navrotsky, J. Chem. Phys. **123**, 194509 (2005).

[7] W. Zheng and S. L. Simon, J. Chem. Phys. **127**, 194501 (2007).

[8] Y. Xia, G. Dosseh, D. Morineau, and C. Alba-Simionesco, J. Phys. Chem. B **110**, 19735 (2006).

[9] D. Morineau, Y. Xia, and C. Alba-Simionesco, J. Chem. Phys., **117**, 8966-8972 (2002).

[10] A. Schonhals, H. Goering, C. Schick, B. Frick, and R. Zorn, Eur. Phys. J. E **12**, 173–8 (2003).

[11] J. Schüller, Y. B. Mel'nichenko, R. Richert and E. W. Fischer, Phys. Rev. Lett., **73**, 2224-2227 (1994).

[12] Yu. B. Mel'nichenko, J. Schüller, R. Richert, and B. Ewen, and, C.-K. Loong, J. Chem. Phys, **103**, 2016-2024 (1995).

[13] P. Pissis, A. Kyritsis, G. Barut, R. Pelster, and G. Nimtz, J. of Non-Cryst. Solids, **235-237**, 444-449 (1998).

[14] C. Le Quellec, G. Dosseh, F. Audonnet, N. Brodie-Linder, C. Alba-Simionesco, W. Häussler, and B. Frick, Eur. Phys. J. S. T., **141**, 11-18 (2007)

[15] R. Zorn, M. Mayorova, D. Richter, and B. Frick, Soft Matter, **4**, 522–533 (2008).

[16] M. Arndt, R. Stannarius, H. Groothues, E. Hempel, and F. Kremer, Phys. Rev. Lett., **79** 2077 (1997).

[17] A. J. Moreno, J. Colmenero, A. Alegria, C. Alba-Simionesco G. Dosseh, D. Morineau, and B. Frick, Eur. Phys. J. E **12**, 43 (2003).

[18] C. Alba-Simionesco, G. Dumont, B. Frick, B. Geil, D. Morineau, V. Teboul, and Y. Xia, Eur. Phys. J. E **12**, 19-28 (2003).

[19] R. Zorn, L. Hartmann, B. Frick, D. Richter and F. Kremer, J. of Non Crys Solids **307–310,** 547 (2002).

[20] G. Dosseh, C. Le Quellec, N. Brodie-Linder, C. Alba-Simionesco, W. Haeussler, and P. Levitz, J. of Non-Crys. Solids, **352**, 4964 (2006).

[21] P. Scheidler, W. Kob, and K. Binder, Eur. Phys. J. E **12**, 5 (2003).

[22] H. Sillescu, J. of Non-Cryst. Solids **243**, 81 (1999).

[23] D. Morineau and C. Alba-Simionesco, J. Chem. Phys. **118**, 9389 (2003).

[24] R. Guégan, D. Morineau, and C. Alba-Simionesco, Chem. Phys. **317**, 236-244 (2005).

[25] D. Morineau, R. Guégan, Y. Xia, and C. Alba-Simionesco, J. Chem. Phys. **121**, 1466 (2004).

[26] D. Kilburn, P. E. Sokol, V. García Sakai, and M. A. Alam, Appl. Phys. Lett. **92**, 033109 (2008).

[27] G. Adam and J. H. Gibbs, J. Chem. Phys. **43**, 139 (1965).

[28] G. Tarjus, D. Kivelson, and P. Viot, J. Phys. Condens. Matter **12**, 6497 (2000).

[29] T. Bellini, L. Radzihovsky, J. Toner, and N. A. Clark, Science **294**, 1074 (2001).

[30] R. Guégan, D. Morineau, C. Loverdo, W. Béziel, and M. Guendouz, Phys. Rev. E **73**, 011707 (2006); A. V. Kityk, M. Wolf, K. Knorr, D. Morineau, R. Lefort, and P. Huber, Phys. Rev. Lett., **101**, 187801. (2008).

[31] R. Guégan, D. Morineau, R. Lefort, A. Moréac, W. Béziel, M. Guendouz, J.-M. Zanotti, and B. Frick, J. Chem. Phys. **126**, 1064902 (2007); R. Lefort, D. Morineau, R. Guégan, M. Guendouz, J.-M. Zanotti, and B. Frick, Phys. Rev. E **78**, 040701(R) (2008).

[32] S. Magazù, F. Migliardo and M. T. F. Telling, J. Phys. Chem. B, **110**, 1020 (2006).

[33] G. Caliskan, D. Mechtani, S. Azzam, J. H. Roh, A. Kisliuk, M. T. Cicerone, S. Lin-Gibson, I. Peral, and A. P. Sokolov, J. Chem. Phys. **121**, 1978 (2004).

[34] J. M. Gordon, G. B. Rouse, J. H. Gibbs, and W. M. Risen Jr., J. Chem. Phys. **66**, 4971 (1966).

[35] V. Caron, J.F. Willart, F. Danede, and M. Descamps, Solid State Comm., **144**, 288 (2007) ; J.F. Willart, N. Descamps, V. Caron, F. Capet, F. Danède, and M. Descamps, Solid State Comm.,**138,** 194-199 (2006).

[36] F. Sussich, R. Urbani, F. Princivalle, and A. Cesaro, J. Am. Chem. Soc., **120**, , 7893-7899 (1998)

[37] D.P. Miller, J.J. de Pablo, J. Phys. Chem. B **104**, 8876 (2000).

[38] R. Busselez, PhD thesis from the University of Rennes 1 (2008); unpublished results.

[39] K. Takeda, K. Murata, and S. Yamashita, J. Phys. Chem. B, **103**, 3457 (1999).

[40] S.S.N. Murthy, and G. Singh, Thermochim. Acta **469**, 116 (2008).

[41] T.E. Dirama, J. E. Curtis, G. A. Carri, and A. P. Sokolov J. Chem. Phys **124**, 34901 (2006) ; T. E. Dirama, G. A. Carri, and A.P. Sokolov. ibid **122**, 244910 (2005); T.E.Dirama G. A. Carri, and A. P. Sokolov ibid **122**, 114505 (2005).

[42] M. T. Cicerone and C. L. Soles, Bioph. J. **86**, 3836 (2004).

[43] M. Guendouz, N. Pedrono, R. Etesse, P. Joubert, J.-F. Bardeau, A. Bulou, and M. Kloul, Phys. Status Solidi A **197**, 414 (2003).

[44] V. Lehmann, R. Stengl, and A. Luigart, Mater. Sci. Eng. B **69**, 11 (2000).

[45] P. Kumar and P. Huber, J. of Nanomat. 2007, 89718 (2007).

[46] D. Wallacher, N. Künzner, D. Kovalev, N. Korr, and K. Knorr, Phys. Rev. Lett. **92**, 195704 (2004).

[47] P. Kumar, T. Hofmann, K. Knorr, P. Huber, P. Scheib and P. Lemmens, J. of Appl. Phys., **103**, 024303 (2008).

[48] B. Frick, B. Farago, in: P. Sabattier (Ed.), Scattering, vol. 2, Academic, 1209, (2002).

[49] L.L.B. is Laboratory Léon Brillouin, CNRS-CEA, Saclay France.

[50] M. Bée, Quasielastic Neutron Scattering, Adam Hilger, Bristol, (1988).

[51] J. Wuttke, W. Petry, G. Coddens and F. Fujara, Phys. Rev. E **52**, 4026 (1995).

[52] O. Sobolev, A. Novikov, and J. Pieper, Chem. Phys., **334**, 36 (2007).

[53] K. Schröter and E. Donth, J. Chem. Phys. **113**, 9101 (2000).

[54] J. Wuttke, I. Chang, O. G. Randl, F. Fujara, and W. Petry, Phys. Rev. E **54**, 5364 (1996).

[55] D. Morineau and C. Alba-Simionesco, J. Chem. Phys. **109**, 8494-8503 (1998).

[56] A. Lerbret, P. Bordat, F. Affouard, M. Descamps, and F. Migliardo, J. Phys. Chem. B **109**, 11046 (2005).

[57] F. He, L.-M. Wang, and R. Richert, Phys. Rev. B **71**, 144205 (2005).

[58] P. Scheidler, W. Kob and K. Binder, Europhys. Lett. **59**, 701 (2002).

[59] C. J. Ellison and J. M. Torkelson, Nature Mater. **2**, 695 (2003).

[60] O. Sobolev, A. Novikov, and J. Pieper, Chem. Phys. **334**, 36 (2007).

[61] A. Arbe, J. Colmenero, F. Alvarez, M. Monkenbusch, D. Richter, B. Farago, and B. Frick, Phys. Rev. E **67**, 051802 (2003).

[62] J. Sacristan, F. Alvarez, and J. Colmenero, Europhys. Lett. **80**, 38001 (2007).

[63] A. Arbe, J. Colmenero, M. Monkenbusch, and D. Richter, Phys. Rev. Lett. **81**, 590 (1998).

[64] J.-M. Zanotti, M.-C. Bellissent-Funel, and S.-H. Chen, Phys. Rev. E. **59**, 3084 (1999).

[65] B. Chen, E. E. Sigmund, and W. P. Halperin, Phys. Rev. Lett. **96,** 145502 (2006).

[66] S. Magazù, G. Maisano, F. Migliardo, and A. Benedetto, Phys. Rev. E **77**, 061802 (2008); S. Magazu, G. Maisano, F. Migliardo, G.Galli, A. Benedetto, D. Morineau, F. Affouard, and M. Descamps, J. Chem. Phys., **129**, 155103 (2008).

[67] B. Frick, C. Alba-Simionesco, G. Dosseh, C. Le Quellec, A.J. Moreno, J. Colmenero, A. Schönhals, R. Zorn, K. Chrissopoulou, S.H. Anastasiadis, K. Dalnoki-Veress, J.of Non-Cryst. Solids 2657, **351** (2005).

[68] K. L. Ngai, Philos. Mag. **84**, 1341 (2004).

[69] C. L. Soles, J. F. Douglas, W. Wu, H. Peng, and D. W. Gidley, Macromolecules **37**, 2890 (2004)

[70] N. Ulbrig, Diploma arbeit from the University of Saarbrücken, (2007).

[71] R. Zondervan, F. Kulzer, G. C. G. Berkhout, and M. Orrit, Proc. Natl. Acad. Sci. USA **104**, 12628 (2007).

[72] R. Zondervan, T. Xia, H. van der Meer, C. Storm, F. Kulzer, W. van Saarloos, and M. Orrit, Proc. Natl. Acad. Sci. USA **105**, 4993 (2008).

[73] D. Sharma and G. S. Iannacchione, J. Chem. Phys. **126**, 094503 (2007).

**Table 1**

Average diffusion coefficient and Kohlrausch stretching exponent obtained from the incoherent quasielastic neutron backscattering function at a temperature $T = 310$ K.

|  | Bulk glycerol | Confined glycerol | Bulk trehalose-glycerol | Confined trehalose-glycerol |
|---|---|---|---|---|
| $\beta_K$ | 0.6 | 0.5 | 0.5 | 0.4 |
| $<D>$ $(10^{-8} cm^2 \cdot s^{-1})$ | 5.0 | 7.6 | 0.52 | 0.77 |

**FIGURE CAPTIONS**

**FIG. 1:** Scanning electron micrographs of the porous silicon film. (a) top view at low magnification showing the 30 μm thick porous layer attached to the silicon substrate and (b) side view at higher magnification.

**FIG. 2:** Adsorption/desorption isotherms of nitrogen at 77 K in porous silicon films.

**FIG. 3:** Incoherent quasielastic spectra of glycerol and the glycerol-trehalose solution in bulk and confined in porous silicon layers measured at $T$=310 K on the high resolution backscattering spectrometer IN16. The intensity is corrected for empty sample contribution and normalised to maximum intensity. The displayed curves have been scaled to one at the maximum intensity for clarity. The resolution function of the apparatus is displayed as a filled gray shape. The solid lines are best fits of the data using a stretched exponential (see text). The graphs correspond to three different values of the transfer of momentum, (a) $Q$=0.54 Å$^{-1}$, (b) $Q$=0.96 Å$^{-1}$ and (c) $Q$=1.33 Å$^{-1}$.

**FIG. 4:** Average relaxation time $\langle \tau \rangle$ as a function of the transfer of momentum $Q$ obtained from the incoherent quasielastic spectra at $T$ =310 K for pure glycerol and the glycerol-trehalose solution in bulk and confined in porous silicon. Dashed lines superimposed on bulk glycerol data emphasize the cross-over between two different power law variations.

**FIG. 5:** Fixed window elastic scattering intensity of glycerol and the glycerol-trehalose solution in bulk and confined in porous silicon layer obtained during a scan on cooling for transfer of momentum $Q$=1.33 Å$^{-1}$. The intensity is corrected for empty sample contribution and normalised at the lowest temperature.

**FIG. 6:** Mean square displacement of glycerol and the glycerol-trehalose solution in bulk and confined in porous silicon layer obtained from incoherent neutron backscattering on cooling. Inset : magnification of the temperature region of the glass transition.

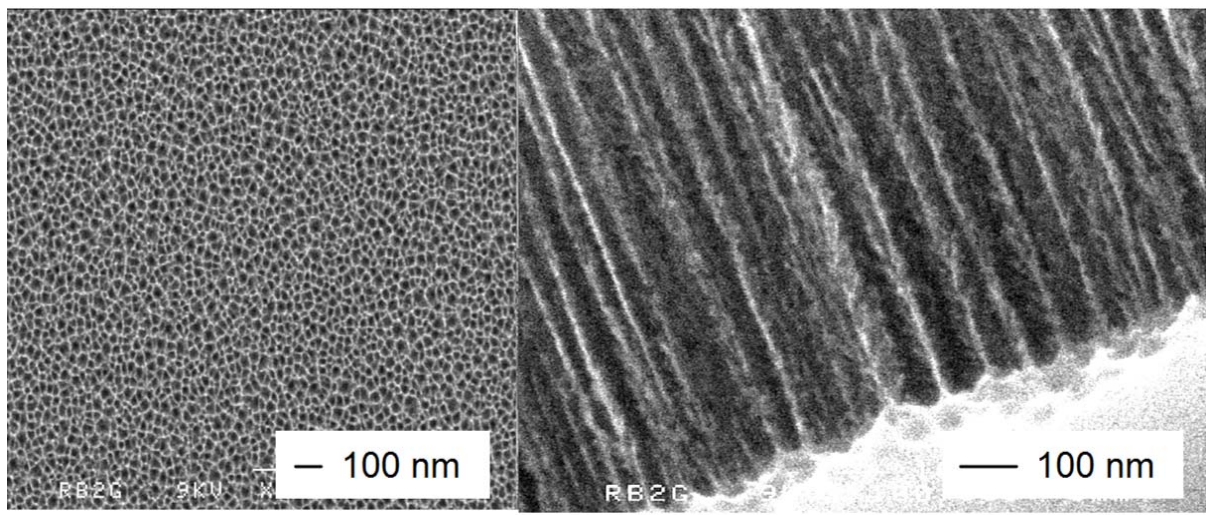

Figure 1
**Molecular dynamics of glycerol and glycerol-trehalose bioprotectant solution nanoconfined in porous silicon.**

R. Busselez et al.

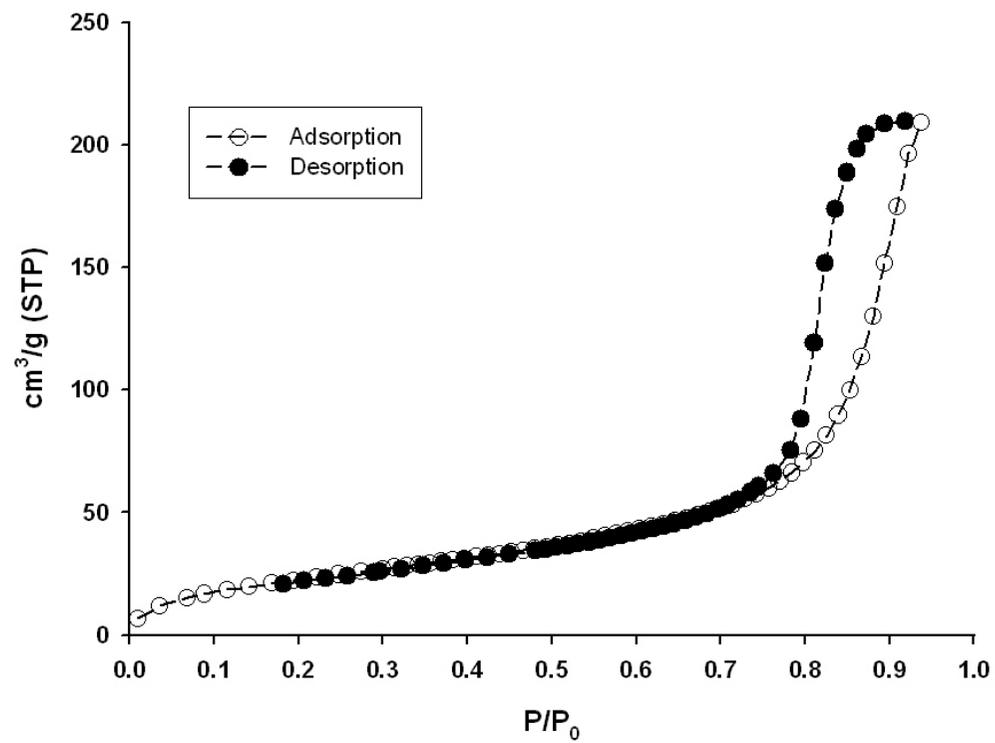

Figure 2
**Molecular dynamics of glycerol and glycerol-trehalose bioprotectant solution nanoconfined in porous silicon.**

R. Busselez et al.

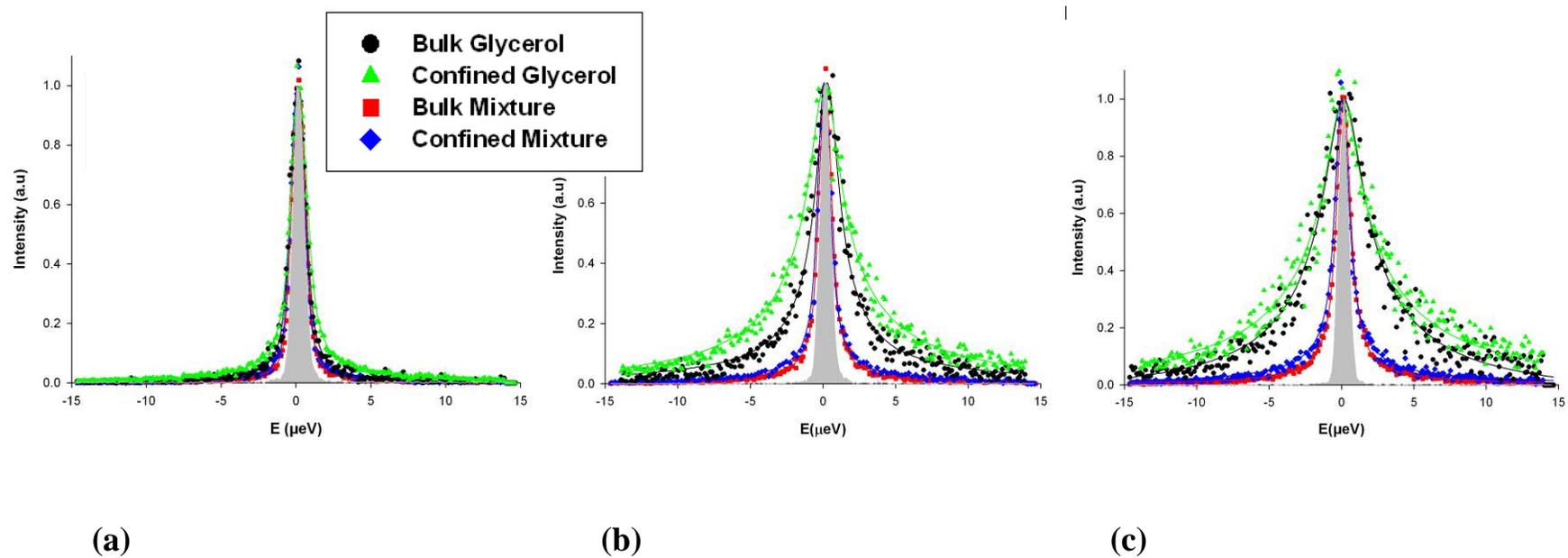

**(a)** **(b)** **(c)**

Figure 3

**Molecular dynamics of glycerol and glycerol-trehalose bioprotectant solution nanoconfined in porous silicon.**

R. Busselez et al.

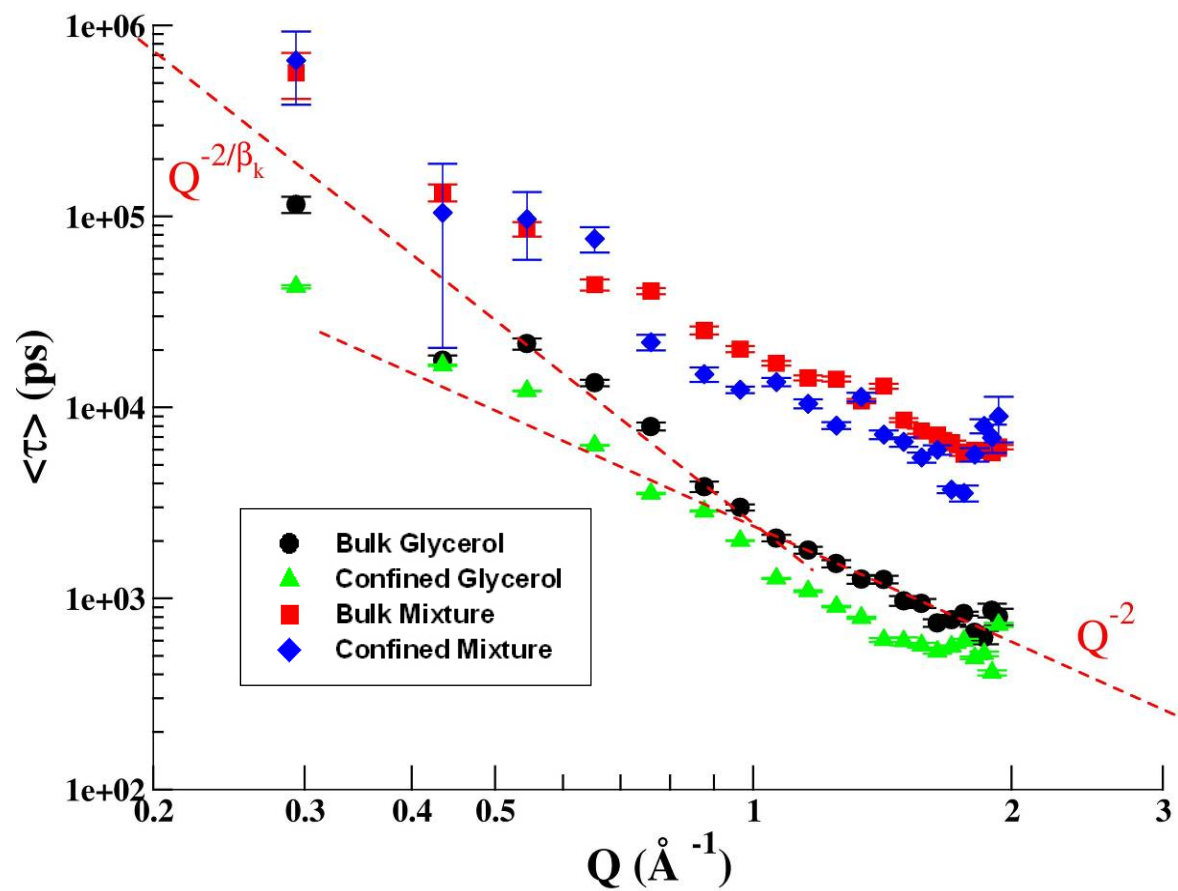

Figure 4
**Molecular dynamics of glycerol and glycerol-trehalose bioprotectant solution nanoconfined in porous silicon.**

R. Busselez et al.

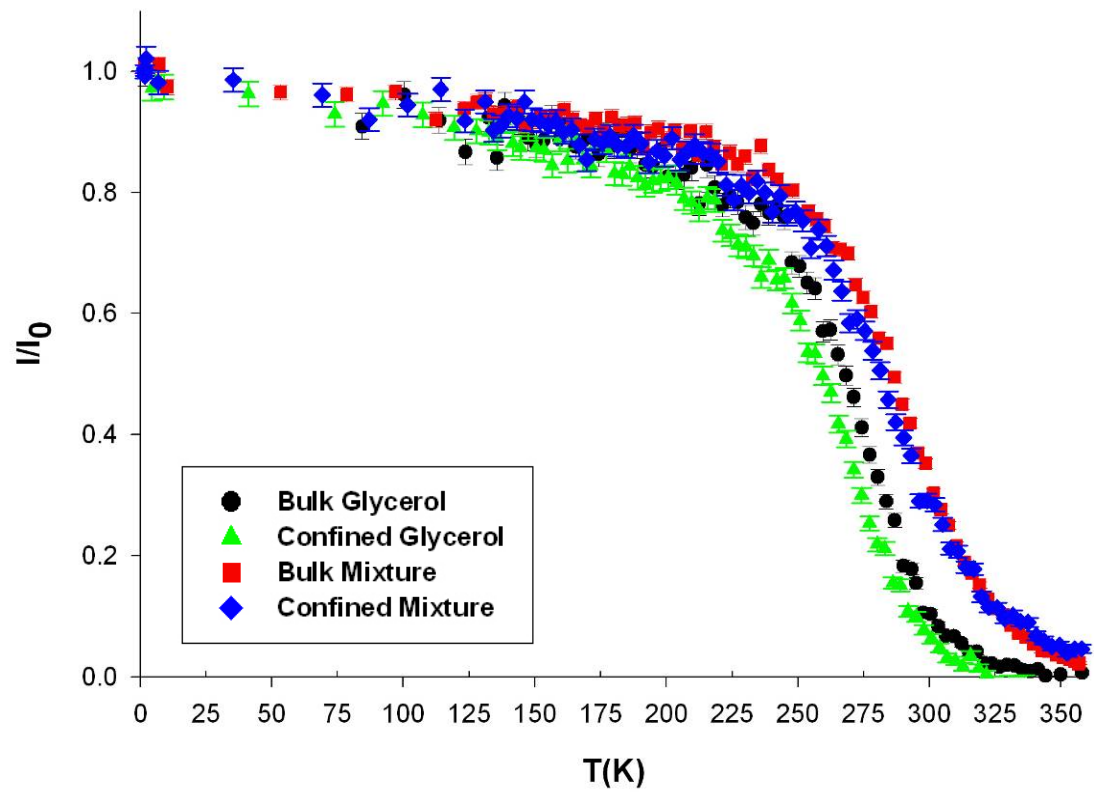

Figure 5
**Molecular dynamics of glycerol and glycerol-trehalose bioprotectant solution nanoconfined in porous silicon.**
R. Busselez et al.

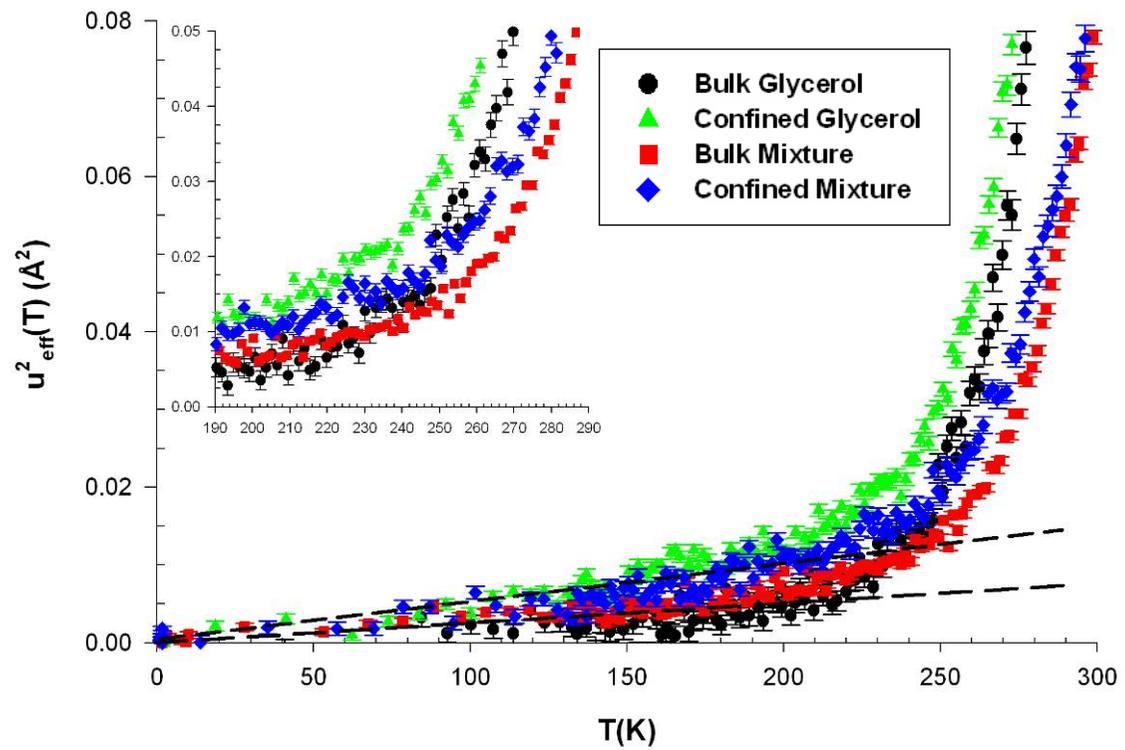

Figure 6
**Molecular dynamics of glycerol and glycerol-trehalose bioprotectant solution nanoconfined in porous silicon.**

R. Busselez et al.